\documentclass[conference]{IEEEtran}
\IEEEoverridecommandlockouts
\usepackage{cite}
\usepackage{amsmath,amssymb,amsfonts}
\usepackage{algorithmic}
\usepackage{graphicx}
\usepackage{textcomp}
\usepackage{xcolor}
\usepackage{hyperref}

\usepackage{url}
\usepackage{booktabs}       % professional-quality tables
\usepackage{multirow}
\usepackage{amssymb}
\usepackage{pifont}
\usepackage{arydshln}
\usepackage{pgfplots}
\pgfplotsset{compat=1.18} 
\usepackage{dsfont}

\def\BibTeX{{\rm B\kern-.05em{\sc i\kern-.025em b}\kern-.08em
    T\kern-.1667em\lower.7ex\hbox{E}\kern-.125emX}}

\begin{document}

\title{Naturalistic Music Decoding from EEG Data via Latent Diffusion Models% {\footnotesize \textsuperscript{*}Note: Sub-titles are not captured in Xplore and
% should not be used}
\thanks{E.P. and E.R. were partially supported by the ERC grant no. 802554 (SPECGEO), PRIN 2020 project no. 2020TA3K9N (LEGO.AI), and PNRR MUR project no. PE0000013-FAIR. E.P. and L.C. were partially supported by PRIN 2022 project no. 2022AL45R2 (EYE-FI.AI, CUP H53D2300350-0001).}
}

\author{
    \IEEEauthorblockN{1\textsuperscript{st} Emilian Postolache*\thanks{*Work partially done during an internship at Sony CSL Tokyo.}}
    \IEEEauthorblockA{\textit{DAIS} \\
    \textit{Ca' Foscari University of Venice} \\
    Venice, Italy \\ emilian.postolache@unive.it 
    }
    \and
     \IEEEauthorblockN{2\textsuperscript{nd} Natalia Polouliakh}
    \IEEEauthorblockA{\textit{Sony CSL} \\
    Tokyo, Japan \\
    nata@csl.sony.co.jp
    }
    \and
    \IEEEauthorblockN{3\textsuperscript{rd} Hiroaki Kitano}
    \IEEEauthorblockA{\textit{Sony CSL}   \\
    Tokyo, Japan  \\ h.kitano@csl.sony.co.jp
    }
    \and
    \IEEEauthorblockN{4\textsuperscript{th} Akima Connelly}
    \IEEEauthorblockA{\textit{Dept. of TSE}
    \\ \textit{Tokyo Institute of Technology}  \\ Tokyo, Japan \\ 
    connelly.a.aa@m.titech.ac.jp}   
    \and
    \IEEEauthorblockN{5\textsuperscript{th} Emanuele Rodol\`a}
    \IEEEauthorblockA{\textit{Dept. of Computer Science} \\
    \textit{Sapienza University of Rome} \\
    Rome, Italy \\ rodola@di.uniroma1.it
    } 
    \and 
    \IEEEauthorblockN{6\textsuperscript{th} Luca Cosmo}
    \IEEEauthorblockA{\textit{DAIS} \\
    \textit{Ca' Foscari University of Venice} \\
    Venice, Italy \\
    luca.cosmo@unive.it
    }
    \and
    \IEEEauthorblockN{7\textsuperscript{th} Taketo Akama}
    \IEEEauthorblockA{\textit{Sony CSL} \\
    Tokyo, Japan \\ taketo.akama@sony.com
    } 
 
}

% \author{
% \begin{tabular}{@{}c@{}}
% Emilian Postolache$^{1*}$ \thanks{$^*$Work partially done during an internship at Sony CSL Tokyo.}
% \qquad Natalia Polouliakh$^{2}$
% \qquad Hiroaki Kitano$^{2}$
% \qquad Akima Connelly$^{3}$\\
% \qquad Emanuele Rodol\`a$^{4}$
% \qquad Luca Cosmo\`a$^{1}$ 
% \qquad Taketo Akama$^{2}$
% \end{tabular}
% \IEEEauthorblockA{ \\
% \\
% $^1$DAIS, Ca' Foscari University of Venice, Italy \quad $^2$Sony CSL, Tokyo, Japan \\
% $^3$Tokyo Institute of Technology, Japan \quad
% $^4$Sapienza University of Rome, Italy \\
% }
% }

\maketitle

\begin{abstract}
In this article, we explore the potential of using latent diffusion models, a family of powerful generative models, for the task of reconstructing naturalistic music from electroencephalogram (EEG) recordings. Unlike simpler music with limited timbres, such as MIDI-generated tunes or monophonic pieces, the focus here is on intricate music featuring a diverse array of instruments, voices, and effects, rich in harmonics and timbre. This study represents an initial foray into achieving general music reconstruction of high-quality using non-invasive EEG data, employing an end-to-end training approach directly on raw data without the need for manual pre-processing and channel selection. We train our models on the public NMED-T dataset and perform quantitative evaluation proposing neural embedding-based metrics. Our work contributes to the ongoing research in neural decoding and brain-computer interfaces, offering insights into the feasibility of using EEG data for complex auditory information reconstruction.
\end{abstract}


\section{Introduction}
\label{sec:intro}

In this paper, we focus on a method that decodes musical information from brain waves captured by electroencephalography (EEG). Reconstruction of musical information conditioned on functional magnetic resonance imaging (fMRI) has been recently proposed in \cite{denk2023brain2music} by mapping the signals to MuLan \cite{huang2022mulan} embeddings and generating with the MusicLM \cite{agostinelli2023musiclm} autoregressive model. Another work \cite{bellier2023music} performs realistic music reconstruction of unseen chunks of a song using electrocorticography (ECoG). The downside of fMRI and, especially, of ECoG, is the difficulty of acquiring data from subjects, since fMRI scanners are unfeasible for real-time capture while ECoG is invasive, placing electrodes inside the subject's skull. In contrast, EEG is non-invasive and can be recorded using wearable devices \cite{apple2023biosignal}.

\begin{figure}[t!]
 \centerline{
 \includegraphics[width=0.85\columnwidth]{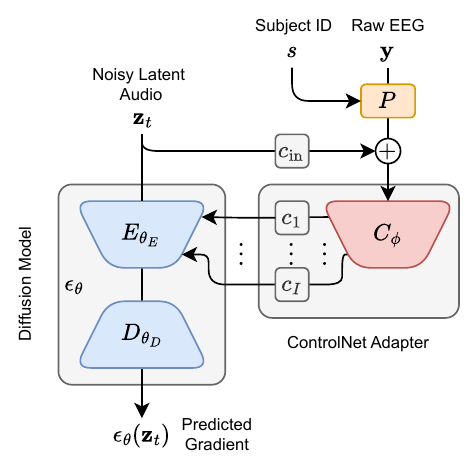}}
 \caption{\textbf{Illustration of proposed method.} We use ControlNet for conditioning a diffusion model on EEG data, in order to decode high-quality naturalistic music.}
 \label{fig:overview}
\end{figure}

To our knowledge, \cite{daly2023neural} is the only work that is able to decode music from EEG signals alone (although with less success than when using a hybrid EEG-fMRI representation). At the same time, the method is limited by  \textit{(i)} the necessity of manually processing and filtering the raw data for successful training of the neural network decoder (a BLSTM \cite{hochreiter1997long}) and \textit{(ii)} the focus on mono-timbral piano music which does not exhibit the variety of realistic music. In the vein of \cite{defossez2023decoding}, which proposes a contrastive method for mapping speech to text without manually processing the dataset, we propose a method which is agnostic to manual pre-processing, based on denosing diffusion \cite{song2019generative, ho2020denoising}. We also target the more challenging scenario of naturalistic music decoding (i.e., complex music containing multiple stems, featuring rich timbre and harmony, as opposed to mono-timbral or MIDI generated music).

Diffusion models \cite{song2019generative, ho2020denoising} have proven to be a surprisingly powerful and flexible class of generative models, being applicable to any type of continuous data, such as images \cite{dhariwal2021diffusion, rombach2022high} and audio \cite{kong2020diffwave, chen2021wavegrad}. In the audio domain, in addition to general audio synthesis \cite{pascual2023full, liu2023audioldm}, various models have been specialized in different tasks ranging from the synthesis of Foley sound \cite{comunita2024syncfusion, chung2024t, lee2024video}, source separation \cite{scheibler2023diffusion, lutati2024separate} and super-resolution \cite{yu2023conditioning}. Musical generation in a continuous domain (i.e., where the representation is not discrete like in MIDI or sheet music), both of a final track \cite{schneider2023mo, evans2024fast} or at the sub-stem level \cite{mariani2023multi, han2023instructme, postolache2024generalized, parker2024stemgen}, is particularly successful when approached with diffusion models, despite the intrinsic difficulty of capturing complex instrumental timbres and the coherence between different stems \cite{ciranni2024cocola}.

\begin{figure}[t!]
 \centerline{
 \includegraphics[width=1.0\columnwidth]{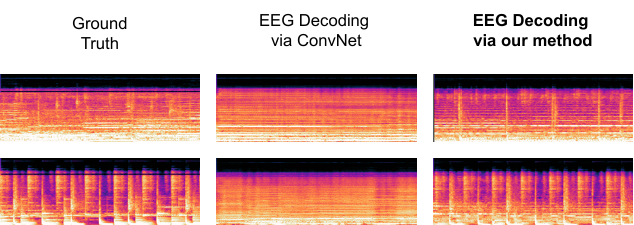}}
 \caption{\textbf{Qualitative results of our method.} On the left ground truth, musical chunks. In the middle, reconstructions obtained via a baseline ConvNet. On the right, decodings obtained by our method. Notice how our method better matches the real tracks.}
 \label{fig:qualitative}
\end{figure}

Recent works \cite{bai2023dreamdiffusion, lan2023seeing} based on Stable Diffusion \cite{rombach2022high} have enabled the synthesis of high-quality images reconstructed from EEG recordings (for a more informative survey about the growing literature on multi-modal brainwave decoding, the reader can consult \cite{mai2023brain}). In this article, we explore the use of ControlNet \cite{zhang2023adding} as a mechanism for controlling AudioLDM2 \cite{liu2023audioldm}, a large-scale pre-trained diffusion model on audio, using EEG data. While such a mechanism has been explored in the EEG to image literature with the CMVDM model \cite{zeng2024controllable}, to our knowledge we are the first work that explores ControlNet for EEG to audio decoding (music in particular), performing such a task \textit{(i)} without manual pre-processing/filtering and \textit{(ii)} leveraging a naturalistic music dataset \cite{Losorelli2017NMEDTAT}.

In Section \ref{sec:background} we review diffusion models and ControlNet. In Section \ref{sec:method} we describe how we apply such an architecture to our setting. In the following, we describe the dataset and experimental setup in Section \ref{sec:setup}, present the results in Section \ref{sec:results}, and conclude the paper in Section \ref{sec:conclusions}.

\begin{figure*}[t!]
 \centerline{
 \includegraphics[width=2.0\columnwidth]{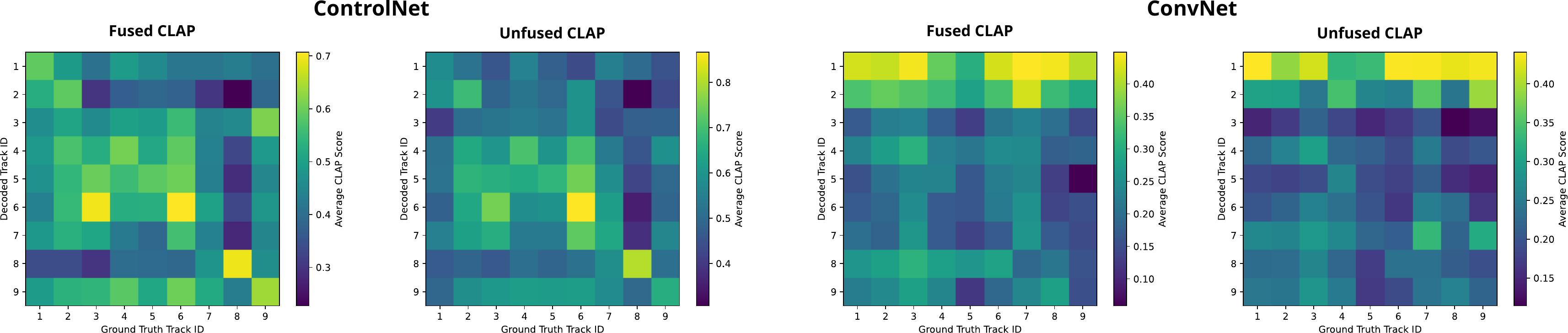}}
 \caption{\textbf{Cross-CLAP scores between decoded and ground truth tracks.} Left: Our method (ControlNet-2). Right: Baseline convolutional network. Notice how the matrices are closer to diagonal with our method, indicating higher correlation (as measured by CLAP score) between decoded and ground truth tracks.}
 \label{fig:classification}
\end{figure*}

\section{Background}
\label{sec:background}

\subsection{Latent Diffusion Models}
\label{subsec:diffusion}
In the following, we review latent diffusion models \cite{rombach2022high, vahdat2021score}, on which we base our methodology. Given a VAE \cite{kingma2014auto} mapping an audio signal (e.g., waveform or mel-spectrogram) $\mathbf{x}$ to a latent representation $\mathbf{z}$ via a stochastic encoder  $E^\text{VAE}$, i.e., $\mathbf{z} \sim E^\text{VAE}(\mathbf{x})$, we can define a prior $p_\theta$ via a diffusion model $\epsilon_\theta$. Based on the DDPM formulation \cite{ho2020denoising}, $\epsilon_\theta$ is trained to revert a forward Gaussian process, indexed by a time variable $t$ ranging in $[0, T]$:
\begin{equation}
\label{eq:forward_process}
    \mathbf{z}_t = \sqrt{\bar{\alpha}_t} \mathbf{z} + \sqrt{1 - \bar{\alpha}_t} \boldsymbol{\epsilon}, \quad \boldsymbol{\epsilon} \sim \mathcal{N} (\mathbf{0}, \mathbf{I})\,,
\end{equation}
with $\alpha_t = 1 - \beta_t$, $\bar{\alpha}_t = \prod_{s=1}^{t} \alpha_s$, and $\{ \beta_t \}_{t=1, \dots, T}$ a noise schedule. 
More specifically, $\epsilon_\theta$ is parameterized by a neural network (typically a U-Net \cite{ronneberger2015u}), and estimates the noisy log gradient $\nabla_{\mathbf{z}_t} \log p_\theta(\mathbf{z}_t)$ of the prior $p_\theta$ via denoising score matching \cite{song2019generative}:
\begin{equation}
\label{eq:score_matching}
\min_\theta \mathbb{E}_{\mathbf{z}, \boldsymbol{\epsilon}, t}\left[\Vert \boldsymbol{\epsilon} -  \epsilon_\theta(\mathbf{z}_t, t) \Vert^2_2\right]\,,
\end{equation}
where we compute $\mathbf{z}_t$ from latent data representations $\mathbf{z}$ using Eq. \eqref{eq:forward_process}.

At inference time, we first sample $\hat{\mathbf{z}}_T \sim \mathcal{N}(\mathbf{0}, \mathbf{I})$ (i.e., the prior of the diffusion process following Eq. \eqref{eq:forward_process}), and compute a backward sequence $\{ \hat{\mathbf{z}}_t\}$ integrating $\epsilon_\theta$ via standard samplers such as DDIM \cite{song2020denoising} (which we employ in this paper). The $\hat{\mathbf{z}}_t$ is finally decoded with the VAE decoder $D^\text{VAE}$ resulting in the waveform output  $\hat{\mathbf{x}} = D^\text{VAE}(\hat{\mathbf{z}}_0)$. 

\subsection{ControlNet}

ControlNet \cite{zhang2023adding} is a parameter-efficient fine-tuning method (PEFT) for (U-Net based) diffusion models, which lets us condition the base model on additional inputs $\mathbf{y}$.  Given the U-Net architecture, we can write the diffusion model $\epsilon_\theta$ as:
\begin{equation*}
\epsilon_\theta(\mathbf{z}_t, t) = D_{\theta_D}(B_{\theta_B}(E_{\theta_E}(\mathbf{z}_t, t))\,,
\end{equation*}
with $E_{\theta_{E}}$ an encoder, $B_{\theta_{B}}$ an inner bottleneck and $D_{\theta{D}}$ a decoder mapping data back to the original dimensions. ControlNet defines an \textit{adapter} $C_\phi$:
\begin{equation}
\label{eq:controlnet}
    C_\phi(\mathbf{z}_t, \mathbf{y}, t) = E_\theta(c_{\text{in}}(\mathbf{z}_t) + P(\mathbf{y}), t)\,,
\end{equation}
initialized with the weights of a pre-trained $E_{\theta_E}$, where $P$ is a projection network mapping $\mathbf{y}$ to a tensor having the dimensions of $\mathbf{z}_t$ and $c_\text{in}$ a zero initialized convolution. While the original ControlNet adapter is additionally conditioned on text (following $E_{\theta_E}$), we omit text conditioning given that we do not use such a modality in this work. The ControlNet adapter outputs, for each layer $i = 1, \dots, I$, a feature $C^i_\phi(\mathbf{z}_t, \mathbf{y}, t)$ which is first convolved with a zero-initialized convolution $c_{i}$, then summed to the features of the original U-Net encoder $E^i_\theta(\mathbf{z}_t, t)$, defining:
\begin{equation}
E^i_{\theta_E, \phi}(\mathbf{z}_t, \mathbf{y}, t) =  c_i(C^i_\phi(\mathbf{z}_t, \mathbf{y}, t)) + E^i_{\theta_E}(\mathbf{z}_t, t)\,.
\end{equation}
 The $c_i$ convolutions improve the generalizability of the fine-tuned model and avoid catastrophic forgetting. The full ControlNet network $\epsilon_{\theta, \phi}$ simply substitutes $E_{\theta_E}$ with $E_{\theta_E, \phi}$ and fine-tunes only the $\phi$ weights, keeping the $\theta$ weights frozen.

\section{Method}
\label{sec:method}
\subsection{EEG-Conditioned ControlNet}
In our method, we employ the ControlNet adapter to condition a (pre-trained) diffusion model on raw EEG information $\mathbf{y} \in \mathbb{R}^{F_{\mathbf{y}} \times S_{\mathbf{y}}}$, where $F_{\mathbf{y}}$ is the number of EEG channels and $S_{\mathbf{y}}$ the number of sampled time-steps. In this work, $E^{\text{VAE}}$ is a 2D convolutional encoder mapping mel-spectrograms $\mathbf{x} \in \mathbb{R}^{F_{\mathbf{x}}\times S_{\mathbf{x}}}$ ($F_\mathbf{x}$ is the number of frequencies and $S_\mathbf{x}$ the number of time bins) to the latent representation $\mathbf{z} \in \mathbb{R}^{F_{\mathbf{z}} \times  S_{\mathbf{z}} \times D_{\mathbf{z}}}$, where $F_\mathbf{z}$ and $S_\mathbf{z}$ are the compressed window dimensions and $D_\mathbf{z}$ the latent channels. As shown in Eq. \eqref{eq:controlnet}, we need to map $\mathbf{y}$ with a projector $P$ to a tensor having the dimensions of $\mathbf{z}$ in order to combine the two sources of information in the ControlNet adapter input. We define the projector $P: \mathbb{R}^{F_{\mathbf{y}} \times S_{\mathbf{y}}} \to \mathbb{R}^{F_{\mathbf{z}} \times  S_{\mathbf{z}} \times D_{\mathbf{z}}}$ as a strided 1D conv-net applied over the temporal dimension of $\mathbf{y}$, followed by a reshape operation.

We also experiment with adding a linear layer $L(\mathbf{y}, s)$ dependent on a subject id $s \in \mathbb{N}$, as defined in \cite{defossez2023decoding}. In such a case, the final projector is obtained as the composition $P \circ L$.  The subject layer modulates the input EEG based on the additional subject id information.

As mentioned in Section \ref{sec:intro}, we only apply minimal pre-processing to $\mathbf{y}$. Namely, we apply a robust scaler \cite{defossez2023decoding} followed by standard deviation clamping. Such methods are data-agnostic, and we do not rely on manual filtering.

\subsection{Neural Embedding-Based Metrics}

Given the low resolution in the sampling of EEG data from current sensors (e.g., 1 kHz) compared to that of audio (e.g., 16 kHz), it is difficult to faithfully reconstruct local track details. Yet, through generative methods such as the proposed model, we can reconstruct tracks at a semantic level, capturing details in a mid-global context (such as mood and timbral characteristics). To assess decoding quality in the aforementioned conditions we need to evaluate the outputs on a more compressed time scale, discarding local information. Therefore, in this paper, we propose to compute evaluation metrics over embeddings $\mathbf{e}$ and $\hat{\mathbf{e}}$ obtained from ground truth tracks $\mathbf{x}$ and generated outputs $\hat{\mathbf{x}}$ (conditioned on EEG data $\mathbf{y}$ corresponding to $\mathbf{x}$) using CLAP \cite{elizalde2023clap, wu2023large} (global scale) and EnCodec \cite{defossez2023high} (intermediate scale) encoders.

We propose to use the following metrics for measuring decoding quality: 
\begin{itemize}
\item
\noindent
\textbf{FAD ---} The Fréchet Audio Distance (FAD) \cite{roblek2019fr} computes global perceptual quality over the distribution of the generations (i.e., without comparing individual samples). It is defined as the Wasserstein distance between Gaussians estimated on embedding sets of ground truth data $\{\mathbf{e}_n\}$ and generated data $\{ \hat{\mathbf{e}}_n\}$. We compute the embeddings both with CLAP and EnCodec. Other embedding spaces, such as VGGish \cite{hershey2017cnn}, are not expressive enough for estimating robust metrics \cite{gui2024adapting}.
\item 
\noindent\textbf{Pearson Coefficient ---} Given embeddings of a ground truth track $\mathbf{e}$ and of a generated track $\hat{\mathbf{e}}$, the (sample) Pearson correlation coefficient $r$ is defined as:
\begin{equation}
\label{eq:pearson}
r = \frac{\sum_i{(\mathbf{e}_i - \boldsymbol{\mu})(\hat{\mathbf{e}}_i - \hat{\boldsymbol{\mu}})} 
}{\sqrt{\sum_i{(\mathbf{e}_i - \boldsymbol{\mu}})^2}\sqrt{\sum_i{(\hat{\mathbf{e}}_i - \hat{\boldsymbol{\mu}})^2}}}\,,
\end{equation}
with $\boldsymbol{\mu}, \hat{\boldsymbol{\mu}}$ the means of $\mathbf{e}, \hat{\mathbf{e}}$, respectively. 
With CLAP embeddings, we notice that the means are very close to zero. Adding that such embeddings have unitary norm, Eq. \eqref{eq:pearson} then becomes equivalent to the \textbf{CLAP Score}, the inner product $\left \langle \mathbf{e}, \hat{\mathbf{e}}\right\rangle$.
\item 
\noindent\textbf{MSE ---} With EnCodec embeddings, we also compute the mean square error $\frac{1}{I}\left \Vert \mathbf{e} - \hat{\mathbf{e}} \right\Vert^2_2$ ($I$ is the maximum time-step).
\end{itemize}

To estimate the statistical significance of the metrics (excluding FAD), we follow \cite{daly2023neural} by computing $p$-values using 4000 bootstrapped null hypotheses obtained by randomly shuffling (with replacement) the indices of the reconstructed audio.

\begin{table*}[!t]
\centering
\caption{\textbf{Quantitative results.}}
\label{tab:results}
 \resizebox{\textwidth}{!}{\begin{tabular}{@{}lllllllll@{}}
\toprule
\textbf{}                                                         & \multicolumn{2}{c}{\textbf{FAD $\downarrow$}} & \multicolumn{2}{c}{\textbf{CLAP Score $\uparrow$}} & \multicolumn{2}{c}{\textbf{Pearson Coefficient (EnCodec) $\uparrow$}} & \multicolumn{2}{c}{\textbf{MSE (EnCodec) $\downarrow$}} \\
\textbf{Experiment} & CLAP        & EnCodec       & Test set                 & OOD Track               & Test set                       & OOD Track                      & Test set                 & OOD Track                    \\ \midrule
Conv-2                                                          & 1.09                & 320.78                  & 0.26                           & -                       &  0.26                      & -                              & 3.19                    & -                            \\
Pretrained                                                   & 0.80                & 163.31                  & 0.38                    & 0.38                            & 0.63                      & 0.64                        & 2.35                     & 2.18                         \\
\hdashline ControlNet-2                                                    & 0.38                & 148.59                  & \textbf{0.60} $(p < 0.01)$            & 0.58                  & \textbf{0.65} $(p < 0.05)$                 & \textbf{0.67}              &  \textbf{2.12}                    & \textbf{1.82}               \\
ControlNet-All                                                  & 0.42                & \textbf{147.06}         & 0.59   $(p < 0.01)$                   & 0.39                 & \textbf{0.65} $(p < 0.01)$                         &  0.65                        & 2.15  $(p < 0.05)$                &  1.94                \\
Scratch-2                                                       & 0.43                & 158.24                  & 0.55   $(p < 0.01)$                   & \textbf{0.60}   $(p < 0.05)$           & 0.61                       & 0.63                      & 2.46                     & 2.14                       \\
Scratch-All                                                     & \textbf{0.36}       & 174.65                  & \textbf{0.60}  $(p < 0.01)$                 & 0.50                     &  0.63 $(p < 0.05)$                         & 0.63                        & 2.42                     & 2.21                         \\ \bottomrule
\end{tabular}}
\end{table*}

\section{Experimental Setup}
\label{sec:setup}
\subsection{Dataset}
We train our models on NMED-T (Naturalistic Music EEG Dataset - Tempo) \cite{Losorelli2017NMEDTAT}. The dataset contains the EEG recordings of 20 adult subjects listening, each, to 10 high-quality songs (not synthesized from MIDI or monotimbral). The EEG data is recorded with the Electrical
Geodesics, Inc. (EGI) GES300 system \cite{tucker1993spatial} using 128 channels, and it is sampled at 1 kHz. We resample the associated audio tracks at 16 kHz. We keep the first track (Trigger 21) as a held-out, out-of-distribution (OOD) track. The other 9 songs are divided into train (0-80\%), validation (80-90\%), and test (90-100\%) subchunks. The models are validated with CLAP score during training. Such a choice stems from the difficulty of doing early stopping with the score matching metric in Eq. \eqref{eq:score_matching}. Our minimal pre-processing pipeline involves excluding the face channels (retaining 124 channels), which we center using the mean over the first 1000 time steps and clamp with 20 standard deviations.

\subsection{Implementation Details}

We implement the proposed ControlNet on top of the AudioLDM2 repository. Specifically, we use the \verb|audioldm2-music| checkpoint, publicly available on HuggingFace\footnote{\url{https://huggingface.co/docs/diffusers/api/pipelines/audioldm2}}. For the ControlNet projection, we use a 1D Conv-Net with (256, 512, 1024, 2048) channels and strides (5, 2, 2, 2). The models process chunks of 3.5 seconds. All models are trained with Adam \cite{kingma2014adam} optimizer using a learning rate of $10^{-4}$. During training and inference, we ignore the textual conditioning mechanism of AudioLDM2 using as an input the constant string ``\verb|Pop music|''. For all metrics in Table \ref{tab:results} we use the unfused version of CLAP.

\section{Experiments}
\label{sec:results}
For our experiments, we train two types of models. First, we train the ControlNet with the frozen diffusion model, using both data from subject 2 (ControlNet-2) and all subjects' data (ControlNet-All). At the same time, we want to test how the model performs when training both the ControlNet adapter and the diffusion U-Net from scratch, leading to the Scratch-2 and Scracth-All models. We compare the models both with Conv-2, a simple convolutional network trained on subject 2 data (following the structure of the projector $P$), and the ``Pretrained" diffusion model, generating audio unconditionally (i.e., no EEG conditioning is occurring). When evaluating ``All'' models, we always use subject 2 test data. We showcase qualitative examples in Figure \ref{fig:qualitative}, and on our site\footnote{\url{https://emilianpostolache.com/brainwave}}, and list quantitative results in Table \ref{tab:results}. The OOD Track represents the held-out track (see Section \ref{sec:setup}). We show $p$-values when metrics are statistically significant ($p < 0.05$). 

On all metrics, the proposed method outperforms the regressor convolutional baseline (which was also trained on the OOD track). We observe that on the Pearson Coefficient and the MSE metrics computed with EnCodec embeddings, the pre-trained model, which does not have access to EEG data, scores  similarly to our models. This indicates that EnCodec embeddings are not optimal for evaluation. On Pearson Coefficient, however, the good $p$-values on our models (except on Scratch-2) indicate higher statistical significance w.r.t. the Pretrained model. On CLAP Score we see a higher improvement with respect to the Conv-2 and Pretrained baselines, reaching a 0.60 CLAP Score on ControlNet-2 and Scrach-All. On FAD, our models also show improved performance. With CLAP Score on the test set, the $p$-values always show statistical significance. Notice that while we often obtain higher results on the OOD Track experiments (especially with EnCodec-based metrics), such metrics have high $p$ values implying low statistical significance. In Figure \ref{fig:classification}, we compute matrices of (cross) CLAP scores between chunks of decoded tracks $i$ and ground truth tracks $j$ (in each entry $i, j$ we average the scores between chunks of tracks $i$ and $j$). We use both the fused and unfused versions of CLAP\cite{wu2023large} for evaluation. The matrices are closer to diagonal with our ControlNet-2 model than with the baseline.

\section{Conclusions}
\label{sec:conclusions}
In this paper, we have investigated the use of latent diffusion models for the task of naturalistic music decoding from EEG brainwaves. While we obtain non-negligible decoding results on unknown segments of known tracks, further research is required to improve generalization to distribution shift, both in terms of larger datasets and improved algorithms.

% For bibtex users:
\bibliographystyle{IEEEtran}
\bibliography{refs}

\end{document}